\documentclass[12pt]{article}
\newcommand{\postscript}[2]
 {\setlength{\epsfxsize}{#2\hsize}
   \centerline{\epsfbox{#1}}}                                                    
\input{epsf}

\def\bit{\begin{itemize}}                                                      
\def\eit{\end{itemize}}
 \newcommand{\note}[1]{}

\newcommand{\delhad}{\mbox{$\Delta \alpha_{\rm had}^{(5)}(M_Z)$}} 
\newcommand{\msb}{\mbox{$\overline{\rm{MS}}\ $}}                                
\newcommand{\mt}{\mbox{$m_t$}}                                                  
\newcommand{\mh}{\mbox{$M_H$}}                                                  
\newcommand{\mz}{\mbox{$M_Z$}}                                                  
\newcommand{\mw}{\mbox{$M_W$}}                                                  
                                      
\newcommand{\als}{\mbox{$\alpha_s$}}                                            
\newcommand{\suf}{\mbox{$SU(5)\ $}}

\newcommand{\skipblk}[1]{}                                                      
\def\bqa{\begin{eqnarray}}                                                      
\def\eqa{\end{eqnarray}}                                                        
\newcommand{\ee}{\mbox{$e^+ e^-$}}

\newcommand{\stto}{\mbox{$SU(2)_{L} \x SU(2)_{R} \x U(1)\ $}\ }                      
\newcommand{\sto}{\mbox{$SU(2) \x U(1)\ $}}                                       
                                               
\newcommand{\x}{\mbox{$\times$}}

\newcommand{\sthto}{\mbox{$SU(3) \x SU(2) \x U(1)\ $}}                             
\newcommand{\sinn}{\mbox{$\sin^2\theta_W\,$}}                                   
                                            
\newcommand{\snu}{\mbox{$\stackrel{(-)}{\nu}$}}                                 
\newcommand{\beq}{\begin{equation}}                                             
\newcommand{\eeq}{\end{equation}}

\newcommand{\RA}{\mbox{$\rightarrow$}}

\def\mxth{\mathsurround=0pt }
\def\xversim#1#2{\lower2.pt\vbox{\baselineskip0pt \lineskip-.5pt
  \ialign{$\mxth#1\hfil##\hfil$\crcr#2\crcr\sim\crcr}}}             
\def\simgr{\mathrel{\mathpalette\xversim >}}                                    
\def\simle{\mathrel{\mathpalette\xversim <}}

\begin{document}

\begin{flushright} UPR-0925T \end{flushright}
\vspace{0.4cm}
\centerline{\LARGE Physics Implications of Precision Electroweak}
\centerline{\LARGE Experiments\footnote{Based on invited talks
presented at the LEP Fest 2000, CERN, October 2000, and the 
Alberto Sirlin Symposium, New York University, October 2000.}}
\vspace{0.8cm}
\centerline{\large Paul Langacker}
\centerline{\large Department of Physics and Astronomy }
\centerline{\large  University of Pennsylvania}
\centerline{\large  Philadelphia, PA 19104}
\vspace{0.4cm}
\centerline{\large  \today }
\vspace{0.9cm}
\begin{abstract}
{A brief review is given of precision electroweak physics,
and its implications for establishing the standard model and constraining
the possibilities for new physics at the TeV scale and beyond.}
\end{abstract}

\newpage

\setcounter{page}{1}

\title{Physics Implications of Precision Electroweak Experiments}
\author{Paul Langacker\\
Department of Physics and Astronomy \\
University of Pennsylvania \\
Philadelphia, PA 19104}
\maketitle
\abstract{A brief review is given of precision electroweak physics,
and its implications for establishing the standard model and constraining
the possibilities for new physics at the TeV scale and beyond.}

\section{Historical Perspective}

The weak neutral current (WNC) and the closely related $W$ and $Z$ properties
have always been the primary prediction and test of
electroweak unification~\cite{general}. Following the original discovery of the WNC in 1973,
there were successive generations of more and more precise experiments,
 culminating in the high precision $Z$ pole
experiments at LEP and SLC in the 1990s and the high energy LEP~2.
The overall result is that: 
  \begin{itemize}
   \item The standard model (SM) is correct and unique to first approximation,
  establishing the gauge
    principle as well as the SM gauge group and representations.
   \item The SM is correct at loop level. This confirms the basic principles of
renormalizable gauge theory, allowed the successful prediction of
$m_t$ and
$\alpha_s$, and strongly constrains  $M_H$.
   \item New  physics at the TeV scale is severely constrained, with the
ideas of unification strongly favored over TeV-scale  compositeness.
   \item The precisely measured  gauge couplings are consistent with 
(supersymmetric) gauge unification.
  \end{itemize}

The  verification of electroweak unification
went through a number of phases. The first was the discovery of WNC
processes  in the Gargamelle bubble chamber at CERN and the HPW
experiment at Fermilab in 1973. The WNC was a
critical prediction of \sto unification, which had been
proposed independently by Weinberg and Salam a few years before,
 and recently shown to be
renormalizable.  Charged current (WCC) processes,
as described by the
Fermi theory, and QED were  incorporated into the theory but not predicted.
However, they were improved and justified, in that it became possible
to calculate meaningful radiative corrections to $\mu$ decay and other
processes. The original \sto theory of leptons was successfully
extended to hadronic processes with the introduction of the charm
quark (the Glashow-Iliopoulos-Maiani
(GIM) mechanism), which was observed in 1974.
   The period saw the simultaneous parallel development of quarks and QCD.

The next generation of experiments, during the late 1970's, were
typically of 10\% precision~\cite{kim}. They included purely WNC
processes, such as  $\snu N \RA \snu N$ (elastic),
$\snu N \RA \snu X$ (inelastic), and $\snu_\mu e \RA \snu_\mu e$,
as well as reactor $\bar{\nu}_e e \RA \bar{\nu}_e e$, for which there are both 
WNC and WCC contributions. In addition, the interference of WNC and
photon exchange effects was observed in the polarization 
asymmetry in $e D \RA e X$ at SLAC. This was crucial because it
excluded pure $S$, $P$,  and $T$ neutral current amplitudes (there was a {\it confusion
theorem} that, in the absence of interference,
 any pure $V$ and $A$ amplitude could be mimicked by a
combination of $S$, $P$,  and $T$), as well as some versions of left-right symmetric 
\stto models. The latter
 had been devised to explain early searches for WNC
parity violation in heavy atoms, which (erroneously)  did not observe any effects.

A number of issues had become important for the interpretation of the experiments.
For example, they were sufficiently precise to require
QCD corrections for the analysis of the deep inelastic neutrino scattering.
Furthermore, the complementarity of experiments became increasingly evident:
there were many types of experiments, probing different WNC couplings in
different kinematic ranges. No one experiment was sensitive to all types
of new physics, and in most cases the new physics effects in a given experiment
could be hidden by a shift in the apparent value of the weak angle \sinn
measured in that process. This led to the need for a global analysis:
 the experiments together
provided much more information than individually.
By  examining their consistency and whether they implied
 a common value for \sinn one could test the SM
and  could separately determine the various WNC couplings in a more general analysis
that did not assume the validity of the SM.
Furthermore, a global analysis permitted the uniform and consistent application of
the best possible theoretical treatments of all similar experiments, and
allowed the proper treatment of common experimental and theoretical uncertainties
and their correlations. On the other hand,  combining experimental results
in a global analysis requires a careful evaluation of systematic uncertainties, since
either underestimations or overestimations can  affect the results.

The global analyses made possible the {\em model independent} fits to
WNC data. These allowed almost
arbitrary ($V-A$ neutrino couplings and family-independent couplings were generally
assumed) vector and axial vector effective four-fermion 
$\nu e$, $\nu q$, and $e q$ WNC
interactions, and therefore parametrized the most general 
family-universal gauge theories involving left-handed neutrinos.
The result  was that the  four-fermi interactions 
predicted by the standard model
 are correct to first approximation, eliminating a number of competing gauge models
which predicted completely different  interactions.
Of course, perturbations around these results, due to new physics, were still allowed.
Furthermore, the new tree-level parameter \sinn of the SM was determined to be
\sinn = 0.229 $\pm 0.010$. There were also limits on
 $\rho - 1$, which parametrizes sources of $SU(2)$ breaking 
such as non-standard Higgs fields or fermion-doublet splittings.
 In modern language, this corresponded to a limit $m_t < 290$ GeV.

The third generation of experiments~\cite{amaldi,costa} began in the 1980's, and continues
with WNC experiments even to the present. They were typically of order
1--5\% precision. They included pure weak $\nu N$ and $\nu e$ scattering processes,
as well as
weak-electromagnetic interference processes such as polarized
$e^{\uparrow \downarrow}D$ or $\mu N$, $\ee \RA $ (hadron or charged lepton)
cross sections and asymmetries below the $Z$ pole, and parity-violating
effects in heavy atoms (APV). The APV experiments had resolved the
experimental difficulties in the earlier generation,
and the most precise involved the hydrogen-like $Cs$ atom, for which one could
cleanly calculate the atomic matrix elements needed to interpret the
results. This generation also included the direct observation
and mass measurements
of the $W$ and $Z$, which were discovered 
by UA1 and UA2 at CERN in 1983. The $Z$ mass in particular allowed the
exclusion of contrived imitator models which had the same
four-fermi operators as the SM, but very different $Z$ spectra.

The increased precision required refined theoretical inputs.
Electroweak radiative corrections had to be
included in the theoretical expressions, and with them one had 
to worry about alternative definitions of the renormalized \sinn.
The most precise results (prior to the $Z$-pole era) involved
deep inelastic scattering. This required careful consideration
of the nucleon structure functions and their QCD evolution,
including a reanalysis of all of the deep-inelastic neutrino
experiments using common structure functions~\cite{amaldi}.
The most precise data involved the ratio of WNC/WCC
scattering on heavy targets that were close to isoscalar, for which
most of the strong interaction uncertainties canceled. The
largest residual uncertainty was  $c$ quark production in the WCC
denominator. Fortunately, this could be measured and parametrized
independently
in dimuon production. Finally, considerable
effort was needed in finding useful parametrizations of such 
 classes of new physics and their effects as heavy
$Z'$ bosons, or new fermions with exotic weak interactions (e.g.,
involving right-handed fields in weak doublets, or left-handed singlets).
It was important that the parametrizations apply to broad classes of models,
but nevertheless that they be simple enough that they could be
easily handled in fits.

The implications of these experimental and theoretical efforts were:
\begin{itemize}
\item The SM is correct to first approximation. That is, 
the four-fermion operators for $\nu q$, $\nu e$, and $eq$ were uniquely determined,
in agreement with the standard model, 
and the $W$ and $Z$ masses agreed with the expectations
of the \sto gauge group and canonical Higgs mechanism, eliminating contrived imitators.
      \item Electroweak radiative corrections  were necessary  for the agreement
of theory and experiment.
        \item The weak mixing angle (in the on-shell renormalization scheme) 
was determined to be \sinn = 0.229 $\pm 0.0064$, while consistency of the various
observations, including radiative corrections,  required
$m_t < 200$ GeV.
\item Theoretical uncertainties, especially in the $c$ threshold 
in deep inelastic WCC scattering,
dominated.
        \item The combination of WNC and WCC data uniquely
determined the $SU(2)$ representations of all of the known fermions,
i.e., of the $\nu_e$ and $\nu_\mu$, as well as the $L$ and
$R$ components of the $e, \ \mu, \ \tau, \ d, \, s, \, b, \ u,$ and $c$~\cite{unique}.
In particular,  the left-handed $b$ was  the
lower component of an $SU(2)$ doublet, implying unambiguously that the $t$ quark
had to exist. This was independent of theoretical arguments based on
anomaly cancellation (which could have been evaded in alternative models
involving a vector-like third family), and of
constraints on the $t$ mass from electroweak loops.
        \item The electroweak gauge couplings were
well-determined, allowing a detailed comparison with the gauge
unification predictions of the simplest grand unified theories (GUT).
It was found that ordinary
    \suf was excluded (consistent with the non-observation of proton decay),
but that the supersymmetric extension was allowed, i.e., that the data
was ``consistent with SUSY
GUTS and perhaps even the first harbinger of supersymmetry''~\cite{amaldi}.
        \item There were stringent limits on new physics at the TeV scale, including
additional $Z'$ bosons, exotic fermions (for which both WNC and WCC 
constraints were crucial), exotic Higgs representations,
 leptoquarks, and new four-fermion operators.
\end{itemize}

In parallel with the WNC--$W$--$Z$, program, there were detailed studies
of $\mu$ decay, which uniquely established its $V-A$ character, and
of the rates and other observables in superallowed and neutron $\beta$ 
decay~\cite{general}.
These constrained such new physics as \stto models, exotic fermions,
and models involving scalar exchange. They also allowed precise determination
of elements of the quark mixing (CKM) matrix and of its unitarity,
thus constraining such effects as a fourth family with significant mixings.
The same period witnessed new experimental confirmations of QCD,
as well as experimental and theoretical developments (in the SM and extensions)
in loop-induced processes such as $b \RA s \gamma$ and, recently, the anomalous
magnetic moment of the muon~\cite{marciano}.

\section{The LEP/SLC Era}

The LEP/SLC era greatly improved the precision of the electroweak program.
It allowed the differentiation between non-decoupling extensions to the
SM (such as most forms of dynamical symmetry breaking and other types
of TeV-scale compositeness), which typically predicted several
\% deviations, and decoupling extensions (such as most of the 
parameter space for supersymmetry), for which the deviations are
typically 0.1\%.

The first phase of the LEP/SLC program involved running at the $Z$
pole, $e^+ e^- \rightarrow Z \rightarrow \ell^+ \ell^-, \ \
  q \bar{q},$ and $\nu \bar{\nu}$. During the period 1989-1995 the
four LEP experiments ALEPH, DELPHI, L3, and OPAL at CERN
observed $\sim  2 \times 10^{7} Z's$. The SLD experiment at the SLC at
SLAC observed some $5 \times 10^5$ events. Despite the much lower statistics,
the SLC had the considerable advantage of a highly polarized $e^-$ beam,
with $P_{e^-} \sim$ 75\%. There were quite a few $Z$ pole observables,
including:
 \begin{itemize}
   \item The lineshape: $M_Z, \Gamma_Z,$ and the peak cross section $ \sigma$
   \item The branching ratios for $e^+e^-,\ \mu^+ \mu^-,\ \tau^+ \tau^-, 
\ q \bar{q},\ c \bar{c},\ b \bar{b},$ and $ s \bar{s}$. One could also determine
the invisible width, $\Gamma({\rm inv})$, from which
one can derive the number 
$N_\nu = 2.985
\pm 0.008$ of active (weak doublet) neutrinos with 
       $m_\nu < M_Z/2$, i.e., there are only 3 conventional families with 
light neutrinos. $\Gamma({\rm inv})$ also constrains other invisible
particles, such as light sneutrinos and the light majorons associated with some 
models of neutrino mass.
   \item A number of asymmetries, including forward-backward (FB) asymmetries; 
the $\tau$ polarization, $P_\tau$;  the polarization asymmetry $A_{LR}$ associated
with $P_{e^-}$; and
mixed polarization-FB asymmetries.
    \end{itemize}
The expressions for the observables are summarized in Appendix~\ref{lineshape},
and the experimental values and SM predictions in Table~\ref{zpole}.
These combinations of observables could be used to isolate many
$Z$-fermion couplings, verify lepton family universality,
determine \sinn in numerous ways, and determine or constrain \mt, \als, and \mh.
 LEP and SLC simultaneously carried out other  programs,
most notably studies and tests of QCD, and heavy quark physics.

LEP~2 ran from 1995-2000, with energies gradually increasing from $\sim 140$ to $\sim 208$ GeV.
The principal electroweak results were precise measurements of the $W$ mass, as well
as its width and branching ratios (these were measured independently at the Tevatron);
a measurement of  $e^+ e^- \RA W^+ W^-$ as a function of center of mass (CM)
energy, which tests the cancellations between diagrams that is characteristic
of a renormalizable gauge field theory, or, equivalently, probes the triple
gauge vertices; stringent lower limits on the Higgs mass, and even hints
of an observation at $\sim$ 115 GeV; limits on anomalous quartic gauge vertices;
and searches for supersymmetric or other exotic particles.

In parallel with the LEP/SLC program, there were much more
precise ($< $ 1\%) measurements of atomic parity violation (APV) in cesium at Boulder,
along with the atomic calculations and related measurements needed for the
interpretation; precise new measurements of deep inelastic
scattering by the NuTeV collaboration at Fermilab, with
a sign-selected beam which allowed them to minimize the effects of the $c$ threshold
and reduce uncertainties to around 1\%; and few \% measurements of $\snu_\mu e$ by CHARM II
at CERN. Although the precision of these WNC processes was  lower
than the $Z$ pole measurements, they are still of considerable importance:
the $Z$ pole  experiments are blind to  types of new physics
that do not directly affect the $Z$,
such as a heavy $Z'$ if there is no $Z-Z'$ mixing,  while the WNC experiments are often very
sensitive. During the same period there were important electroweak results 
from CDF and D$\not{\! 0}$ at the Tevatron, most notably a precise value for $M_W$,
competitive with and complementary to the LEP~2 value; a direct measure of \mt,
and direct searches for  $Z'$, $W'$, exotic fermions, and supersymmetric particles.
Many of these non-$Z$ pole results are summarized in Table~\ref{nonzpole}.

\begin{table} \centering
\begin{tabular}{|l|c|c|c|r|}
\hline Quantity & Group(s) & Value & Standard Model & pull \\ 
\hline
$M_Z$ \hspace{14pt}      [GeV]&     LEP     &$ 91.1876 \pm 0.0021 $&$ 91.1874 \pm 0.0021 $&$ 0.1$ \\
$\Gamma_Z$ \hspace{17pt} [GeV]&     LEP     &$  2.4952 \pm 0.0023 $&$  2.4963 \pm 0.0016 $&$-0.5$ \\
{ $\Gamma({\rm had})$\hspace{8pt}[GeV]}&
 LEP  &$  1.7444 \pm 0.0020 $&$  1.7427 \pm 0.0015 $&  ---  \\
{ $\Gamma({\rm inv})$\hspace{11pt}[MeV]}& LEP  &
$499.0    \pm 1.5    $&$501.74   \pm 0.15   $&  ---  \\
{ $\Gamma({\ell^+\ell^-})$ [MeV]}& 
   LEP     &$ 83.984  \pm 0.086  $&$ 84.018  \pm 0.028  $&  ---  \\
$\sigma_{\rm had}$ \hspace{12pt}[nb]&LEP    &$ 41.541  \pm 0.037  $&$ 41.479  \pm 0.014 
$&{ $1.7$} \\
$R_e$                         &     LEP     &$ 20.804  \pm 0.050  $&$ 20.743  \pm 0.018  $&$ 1.2$ \\
$R_\mu$                       &     LEP     &$ 20.785  \pm 0.033  $&$ 20.743  \pm 0.018  $&$ 1.3$ \\
$R_\tau$                      &     LEP     &$ 20.764  \pm 0.045  $&$ 20.788  \pm 0.018  $&$-0.5$ \\
$A_{FB} (e)$                  &     LEP     &$  0.0145 \pm 0.0025 $&$  0.0165 \pm 0.0003 $&$-0.8$ \\
$A_{FB} (\mu)$                &     LEP     &$  0.0169 \pm 0.0013 $&$                    $&$ 0.3$ \\
$A_{FB} (\tau)$               &     LEP     &$  0.0188 \pm 0.0017 $&$                    $&$ 1.4$ \\
\hline
$R_b$                         &  LEP + SLD  &$  0.21653\pm 0.00069$&$  0.21572\pm 0.00015$&$ 1.2$ \\
$R_c$                         &  LEP + SLD  &$  0.1709 \pm 0.0034 $&$  0.1723 \pm 0.0001 $&$-0.4$ \\
$R_{s,d}/R_{(d+u+s)}$         &     OPAL    &$  0.371  \pm 0.023  $&$  0.3592 \pm 0.0001 $&$ 0.5$ \\
$A_{FB} (b)$                  &     LEP     &$  0.0990 \pm 0.0020 $&$  0.1039 \pm 0.0009 $&{ $-2.5$} \\
$A_{FB} (c)$                  &     LEP     &$  0.0689 \pm 0.0035 $&$  0.0743 \pm 0.0007 $&$-1.5$ \\
$A_{FB} (s)$                  &DELPHI,OPAL&$  0.0976 \pm 0.0114 $&$  0.1040 \pm 0.0009 $&$-0.6$ \\
$A_b$                         &     SLD     &$  0.922  \pm 0.023  $&$  0.9348 \pm 0.0001 $&$-0.6$ \\
$A_c$                         &     SLD     &$  0.631  \pm 0.026  $&$  0.6683 \pm 0.0005 $&$-1.4$ \\
$A_s$                         &     SLD     &$  0.82   \pm 0.13   $&$  0.9357 \pm 0.0001 $&$-0.4$ \\
\hline
$A_{LR}$ (hadrons)            &     SLD     &$  0.15138\pm 0.00216$&$  0.1483 \pm 0.0012 $&$ 1.4$ \\
$A_{LR}$ (leptons)            &     SLD     &$  0.1544 \pm 0.0060 $&$                    $&$ 1.0$ \\
$A_\mu$                       &     SLD     &$  0.142  \pm 0.015  $&$                    $&$-0.4$ \\
$A_\tau$                      &     SLD     &$  0.136  \pm 0.015  $&$                    $&$-0.8$ \\
$A_e (Q_{LR})$                &     SLD     &$  0.162  \pm 0.043  $&$                    $&$ 0.3$ \\
$A_\tau ({\cal P}_\tau)$      &     LEP     &$  0.1439 \pm 0.0042 $&$                    $&$-1.0$ \\
$A_e ({\cal P}_\tau)$         &     LEP     &$  0.1498 \pm 0.0048 $&$                    $&$ 0.3$ \\
$\bar{s}_\ell^2 (Q_{FB})$     &     LEP     &$  0.2321 \pm 0.0010 $&$  0.23136\pm 0.00015$&$ 0.7$ \\
\hline
\end{tabular}
\caption{Principal $Z$-pole observables, their experimental values, 
theoretical predictions using the SM parameters from the global best fit~\cite{erler}, and pull
(difference from the prediction divided by the uncertainty).
$\Gamma({\rm had})$, $\Gamma({\rm inv})$, and $\Gamma({\ell^+\ell^-})$ are not independent,
but are included for completeness.}
\label{zpole}
\end{table}

\begin{table} \centering
\begin{tabular}{|l|c|c|c|r|}
\hline Quantity & Group(s) & Value & Standard Model & pull \\ 
\hline
$m_t$\hspace{8pt}[GeV]&Tevatron &$ 174.3    \pm 5.1               $&$ 174.2    \pm 4.4    $&$ 0.0$ \\
$M_W$ [GeV]    &      LEP       &$  80.427  \pm 0.046             $&$  80.394  \pm 0.019  $&$ 0.7$ \\
$M_W$ [GeV]    & Tevatron,UA2 &$  80.451  \pm 0.061             $&$                     $&$ 0.9$ \\
\hline
$R^-$          &     NuTeV      &$   0.2277 \pm 0.0021 \pm 0.0007 $&$   0.2301 \pm 0.0002 $&$-1.1$ \\
$R^\nu$        &     CCFR       &$   0.5820 \pm 0.0027 \pm 0.0031 $&$   0.5834 \pm 0.0004 $&$-0.3$ \\
$R^\nu$        &     CDHS       &$   0.3096 \pm 0.0033 \pm 0.0028 $&$   0.3093 \pm 0.0002 $&$ 0.1$ \\
$R^\nu$        &     CHARM      &$   0.3021 \pm 0.0031 \pm 0.0026 $&$                     $&{ $-1.8$}
\\
$R^{\bar\nu}$  &     CDHS       &$   0.384  \pm 0.016  \pm 0.007  $&$   0.3862 \pm 0.0002 $&$-0.1$ \\
$R^{\bar\nu}$  &     CHARM      &$   0.403  \pm 0.014  \pm 0.007  $&$                     $&$ 1.0$ \\
$R^{\bar\nu}$  &     CDHS 1979  &$   0.365  \pm 0.015  \pm 0.007  $&$   0.3817 \pm 0.0002 $&$-1.0$ \\
\hline
$g_V^{\nu e}$  &     CHARM II   &$  -0.035  \pm 0.017             $&$  -0.0399 \pm 0.0003 $&  ---  \\
$g_V^{\nu e}$  &      all       &$  -0.041  \pm 0.015             $&$                     $&$-0.1$ \\
$g_A^{\nu e}$  &     CHARM II   &$  -0.503  \pm 0.017             $&$  -0.5065 \pm 0.0001 $&  ---  \\
$g_A^{\nu e}$  &      all       &$  -0.507  \pm 0.014             $&$                     $&$ 0.0$ \\
\hline
$Q_W({\rm Cs})$&     Boulder    &$ -72.65   \pm 0.28\pm 0.34      $&$ -73.08   \pm 0.04   $&$ 1.0$ \\
$Q_W({\rm Tl})$&Oxford,Seattle&$-114.8    \pm 1.2 \pm 3.4       $&$-116.6    \pm 0.1    $&$ 0.5$ \\
\hline
${\Gamma (b\rightarrow s\gamma)\over \Gamma (b\rightarrow c e\nu)}$& CLEO 
           &$ 3.26^{+0.75}_{-0.68} \times 10^{-3} $&$ 3.15^{+0.21}_{-0.20}
           \times 10^{-3} $&$ 0.1$ \\
\hline
\end{tabular}
\caption{Recent non-$Z$-pole observables.}
\label{nonzpole}
\end{table}

The LEP and (after initial difficulties) SLC programs were
remarkably successful, achieving  greater precision than
had been anticipated in the planning stages, e.g., due to better 
than expected measurements of the   beam energy
(using a clever resonant depolarization technique) and luminosity.
Credit goes to the individuals who built and operated
the machines and computing systems, and to the experimenters who
built, ran, and analyzed the results from
 the ALEPH, DELPHI, L3, OPAL (LEP) and  SLD (SLC)
detectors. 
The measurement of the $Z$ mass and width at LEP
were so precise that the tidal effects of the moon, the levels of the water table
and Lake Geneva, and electromagnetic effects from trains had to be
taken into account. SLAC had an advantage of a  high and well-measured
$e^-$ polarization. (One regret is that LEP never implemented longitudinal
polarization. The Blondel scheme would have permitted a high, self-calibrated
polarization with its significant advantages.) The program was greatly
enhanced by the efforts of the LEP Electroweak Working Group (LEPEWWG)~\cite{LEPEWWG}, 
which combined the results of
the four LEP experiments, and also those of SLD and some WNC and Tevatron
results, taking proper account of
common systematic and theoretical uncertainties, so that the maximum and most reliable
information could be extracted.

A great deal of supporting theoretical effort was also
essential. This included
the calculation of the needed electromagnetic,
electroweak,  QCD, and mixed radiative corrections
to the predictions of the SM~\cite{bardin,hollik,passarino}. Careful consideration of
the competing definitions of the renormalized \sinn
was needed. 
The principal theoretical uncertainty is the hadronic
contribution \delhad \ to the running of 
$\alpha$ from its precisely known value at low energies~\cite{kinoshita}
to the $Z$-pole~\cite{jegerlehner}, where it is needed to compare
the $Z$ mass with the asymmetries and other observables.
The radiative corrections, renormalization schemes, and
running of $\alpha$ are further discussed in Appendix~\ref{radiativecorr}.
Tremendous theoretical effort went into the development,
testing, and comparison of radiative corrections packages such
as ZFITTER, TOPAZ0, ALIBABA, BHLUMI, and GAPP, as well as other
packages needed for the LEP~2 program. Finally, much
effort went into the study of how various classes of new
physics would modify the observables, and how they could
most efficiently be parametrized.

Global analyses of the $Z$-pole and other data were carried out, e.g., 
by the LEPEWWG~\cite{LEPEWWG} and by Erler and myself for the Particle Data Group 
(PDG)~\cite{erler}, to test the consistency of the SM, determine its
parameters, and search for or constrain new TeV-scale physics.

Although the $Z$-pole program has ended for the time being, there are
prospects for future programs using the Giga-$Z$ option at TESLA or possible
other linear colliders, which might yield a factor $10^2$ more events. This would 
enormously improve the sensitivity~\cite{giga}, but would also require a large
theoretical effort to improve the radiative correction 
calculations~\cite{bardin,passarino}.

During the LEP/SLC era, the $Z$-pole, LEP~2, WNC, and Tevatron experiments
successfully tested the SM at 0.1\% level, including electroweak loops, thus
confirming the gauge principle,
 SM   group, representations, and the basic structure of renormalizable field theory.
The standard model parameters $\sin^2 \theta_W$, $m_t$, and $\alpha_s$
were precisely determined.
In fact, \mt \ was successfully predicted from its indirect loop effects prior
to the direct discovery at the Tevatron, while the indirect value of \als,
mainly from the $Z$-lineshape, agreed with direct determinations.
Similarly, \delhad \ and $ M_H$ were constrained.
The indirect (loop) effects implied $M_H \simle 194$ GeV, while direct
searches at LEP~2 yielded $M_H > 112 $ GeV, with a hint of a signal at 115 GeV.
This range is consistent with, but does not prove, 
the expectations of the supersymmetric
extension of the SM (MSSM), which predicts a light SM-like Higgs for much of
its parameter space. The agreement of the data with the SM imposes
a severe constraint on possible new physics at the TeV scale,
and points  towards decoupling theories (such as most versions of
supersymmetry and unification), which typically lead to 0.1\% effects,
rather than TeV-scale compositeness (e.g., dynamical symmetry breaking
or composite fermions), which usually imply  deviations of several \% (and often
large flavor changing neutral currents). 
Finally, the precisely measured gauge couplings were consistent with the
simplest form of grand unification if the SM is extended to the MSSM.


\section{Global Electroweak Fits}
\label{globalfits}
Global fits allow uniform theoretical treatment and exploit the
fact that the data collectively contain much more information than 
individual experiments. However, they require a careful
consideration of experimental and theoretical systematics and their correlations.
The results here are  from work with Jens Erler,
updated from the electroweak review in the 2000 {\em Review of 
Particle Properties}~\cite{erler}. They incorporate
the full $Z$-pole, WNC (especially important for constraining some types of new physics),
and relevant hadron collider and LEP~2 results. The radiative corrections were
calculated with Erler's new GAPP ({\em Global Analysis of Particle Properties})
program~\cite{GAPP}. GAPP is fully \msb, which minimizes the mixed QCD-EW corrections
and their uncertainties and is a complement to ZFITTER, which is on-shell.
We use a new \delhad \  which is properly correlated with \als~\cite{alhad}.
Our results are in good agreement with the LEPEWWG~\cite{LEPEWWG} up 
to well-understood effects,
such as more extensive
WNC inputs and small differences in higher order terms
and \delhad,  despite the different
renormalization schemes used.

The data are in excellent agreement with the SM predictions. The best fit
values for the SM parameters (as of 10/00) are,
\bqa
           M_H &=& 86^{+48}_{-32} \mbox{ GeV}, \nonumber \\
           m_t &=& 174.2  \pm 4.4  \mbox{ GeV}, \nonumber  \\
      \alpha_s &=& 0.1195 \pm 0.0028, \label{fitresults} \\
   \hat{s}^2_Z &=& 0.23107 \pm 0.00016, \nonumber \\
 \Delta \alpha_{\rm had}^{(5)}(M_Z) &=& 0.02778 \pm 0.00020 \nonumber
\eqa
\bit 

\item 
This fit included the direct (Tevatron) measurements of \mt \ and
the theoretical value of \delhad \ as constraints, but
did not include other determinations of \als \ or the LEP~2 direct limits
on \mh.

\item
The \msb value of \sinn  ($\hat{s}^2_Z$) can be translated
into  other definitions. The effective angle $\bar{s}^2_\ell = 0.23136 \pm 0.00015$
is closely related to $\hat{s}^2_Z$. The larger  uncertainty in the
on-shell $s^2_W = 0.22272 \pm 0.00038$ is due to its (somewhat artificial) dependence 
on \mh \ and \mt.
On the other hand,  the $Z$-mass definition $s^2_{M_Z} =  0.23105 \pm 0.00008$
has no \mh \ or \mt \ dependence, but the uncertainties reemerge
when comparing with other observables.

\item The best fit value $ \Delta \alpha_{\rm had}^{(5)}(M_Z) = 0.02778 \pm 0.00020$
is dominated by the theoretical input constraint $ \Delta \alpha_{\rm had}^{(5)}(M_Z) =
0.02779 \pm 0.00020$. However, $ \Delta \alpha_{\rm had}^{(5)}(M_Z)$ can be determined from the
indirect data alone, i.e., from the relation of $M_Z$ and $M_W$ to the
other observables. The result,
$0.02765 \pm 0.00040$, is in impressive agreement with the theoretical value. 

\item
Similarly, the value $m_t = 174.2  \pm 4.4$ GeV includes the direct Tevatron constraint
$\mt = 174.3 \pm 5.1$. However, one can determine 
$\mt = 174.1^{+9.7}_{-7.6}$ GeV from indirect data (loops) only, in excellent agreement.

\item The value \als $=0.1195 \pm 0.0028$ is consistent with  other 
determinations, e.g., from deep inelastic scattering, hadronic $\tau$
decays,  the charmonium and upsilon spectra, and
jet properties. The current
PDG average (excluding the $Z$ lineshape) is 0.1182 $\pm$ 0.0013.

 \item
The central value of the 
Higgs mass prediction from the fit, \mh $= 86^{+48}_{-32}$ GeV,
is below the direct lower limit from LEP~2 of $\simgr 112$ GeV,
or their candidate events at 115 GeV, but consistent at the $1\sigma$ level. 
Including the direct LEP~2
likelihood function~\cite{erlerhiggs,degrassi}  along with the indirect data, one obtains
 $\mh < 194$ GeV at 95\%. Even though \mh \ only enters the precision
data logarithmically (as opposed to the quadratic \mt \ dependence),
the constraints are significant. They are also fairly
robust to most, but not all, types of new physics.
(The limit on \mh \ disappears if one allows an arbitrarily large negative
$S$ parameter (section~\ref{newphysics}), but most extensions of the
SM yield $S > 0$.)
The predicted range should be compared with the
theoretically expected range in the standard model: 
115 GeV $\simle \mh \simle$ 750 GeV, where the lower (upper) limit
is from vacuum stability (triviality). On the other hand,
the MSSM predicts $\mh \simle 130$ GeV, while the limit increases to
around 150 GeV in extensions of the MSSM.

\item 
The results in (\ref{fitresults}) are consistent with those
of the LEPEWWG. For example, at Osaka Gurtu presented~\cite{gurtu}:
 $\bar{s}^2_\ell = 0.23140 \pm 0.00016$; \
 $\alpha_s = 0.1183 \pm 0.0027$; \
$m_t = 174.3^{+4.4}_{-4.1}$ GeV; and
$M_H =60^{+52}_{-29}$ GeV. These are in excellent agreement,
except for the somewhat lower \mh. The difference is mainly due to their
use of an earlier estimate~\cite{ej}
$\Delta \alpha_{\rm had}^{(5)}(M_Z) =0.02803 \pm 0.00065$. Gurtu reported
that their \mh \ increased to the consistent $88^{+60}_{-37}$ GeV when they
used a newer $\Delta \alpha_{\rm had}^{(5)}(M_Z) = 0.02755 \pm 0.00046$~\cite{Pietrzyk}
 based on the
 new BES-II $e^+ e^-$ data.

\eit

\section{Beyond the standard model}
\label{newphysics}
The standard model (\sthto plus general
relativity), extended to include neutrino mass,
is the correct description of nature to first approximation down 
to $10^{-16}$ cm. However, nobody thinks that the SM is the ultimate description
of nature. It has some 28 free parameters; has a complicated gauge group and
representations; does not explain charge quantization, the fermion families,
or their masses and mixings; has several notorious fine tunings
associated with the Higgs mass, the strong CP parameter, and
the cosmological constant; and does not incorporate quantum gravity.

Many types of possible TeV scale physics are constrained by the
precision data. For example,

\bit
  \item
$S, T,$ and $ U$ parametrize  new physics sources which only affect the gauge
propagators, as well as Higgs triplets, etc.  One expects $T \ne 0$, usually positive and
often of order
unity,  from nondegenerate heavy
fermion or scalar doublets, 
while new chiral fermions (e.g., in extended technicolor (ETC)), lead to $S \ne 0$,
again usually positive and often of order unity.
The current global fit result is~\cite{erler}
 \bqa  S &=& -0.05 \pm 0.11 (-0.09)  \nonumber  \\
T &=& -0.03 \pm 0.13 (+0.10) \label{stu}  \\
U &=& 0.18 \pm 0.14 (+0.01) \nonumber 
\eqa
for $M_H = 115 \ (340)$ GeV. (We use a definition in which $S$, $T$, and $U$ 
are exactly zero in 
the SM.) The value of $S$ would be $2/3\pi$ for a heavy degenerate ordinary
or mirror family, which is therefore excluded at 99.92\%. Equivalently,
the number of families is $N_{\rm fam} = 2.84 \pm 0.30$. This is
complementary to the lineshape result $N_\nu = 2.985 \pm 0.008 $, which only applies for
 $\nu$'s
lighter than $\sim M_Z/2$. $S$ also eliminates many QCD-like ETC models.
$T$ is equivalent to the $\rho_0$ parameter~\cite{general}, which
is defined to be exactly unity in the SM. For $S = U = 0$, one obtains
 $\rho_0  \sim 1 + \alpha T = 1.0004^{+0.0018}_{-0.0011}$, with the SM
fit value for \mh \ increasing to $M_H = 113^{+310}_{-64}$ GeV.

  \item 
Supersymmetry: in the 
decoupling limit, in which the sparticles are heavier than
$ \simgr 200-300$ GeV, there is little effect on the precision
observables, other than that there is necessarily 
  a light SM-like Higgs, consistent with the data. There is little
improvement on the SM fit, and in fact one can somewhat constrain
the supersymmetry breaking parameters~\cite{susy}.

\item Heavy $Z'$ bosons are predicted by many 
grand unified  and string theories. Limits on the $Z'$ mass
are model dependent, but are typically  around $M_{Z'} > 500-800 $ GeV 
from indirect constraints from WNC and  LEP~2 data, with comparable
limits from direct searches at the Tevatron. $Z$-pole data
severely constrains the $Z-Z'$ mixing, typically
 $|\theta_{Z-Z'}| < {\rm few} \times 10^{-3}$.

\item Gauge unification is predicted in GUTs and string theories.
The simplest non-supersymmetric unification is excluded by
the precision data. For the MSSM, and assuming 
no new thresholds between 1 TeV and the unification scale, one
can use the precisely known $\alpha$ and $\hat{s}^2_Z$
to predict $\als = 0.130 \pm 0.010$ and a unification scale 
$M_G \sim 3 \times 10^{16}$ GeV~\cite{polonsky}. The \als \ uncertainties are 
mainly theoretical, from the TeV and GUT thresholds, etc.
\als \ is high compared to the experimental value, but barely consistent 
given the uncertainties. $M_G$ is reasonable for a GUT (and
is consistent with simple seesaw models of neutrino mass),
but is somewhat below the expectations $\sim 5 \times 10^{17}$ GeV of the simplest
perturbative heterotic string models. However, this is only a
10\% effect in the appropriate variable
$\ln M_G$. The new exotic particles often present in such models
(or higher Ka\v c-Moody levels) can easily shift the $\ln M_G$
and \als \ predictions significantly, so the problem is really
why the gauge unification works so well.
It is always possible that the apparent success is accidental
(cf., the discovery of Pluto).
\eit

\section{Conclusions}

\begin{itemize}
 \item The WNC, $Z$, and $W$ are the primary predictions
 and tests of electroweak unification.
   \item The standard model (SM) is correct and unique to first approximation,
establishing the gauge
    principle, group, and representations.
   \item The SM is correct at the loop level,
confirming renormalizable gauge theory, and successfully predicting or constraining
 $m_t$, $\alpha_s$,
 and   $M_H$.
   \item TeV physics is severely constrained, with the ideas of unification 
favored over TeV-scale  compositeness.
   \item The precisely measured gauge couplings are consistent with gauge unification.
\end{itemize}

Alberto Sirlin and his collaborators have pioneered the calculation of
electromagnetic and electroweak radiative corrections to weak
observables, and their importance for testing first the Fermi theory and then
the SM, as well as constraining new physics.
This has included seminal contributions to the radiative corrections
to $\mu$ and $\beta$ decay; the electroweak corrections to WNC and WCC
processes; tests of the unitarity of the CKM quark mixing matrix; 
the on-shell and \msb renormalization
schemes and definitions of the renormalized \sinn;  the precise relations of $Z$ and $W$ pole
observables such as $M_Z$, $M_W$, and $Z$-pole asymmetries;
and the dependence of all of the above on the $t$ and Higgs masses.
Alberto has played a unique and indispensable role in the development
and testing of the standard model and of renormalizable gauge field theories.
My one major collaboration with Alberto~\cite{amaldi}
was one of the most pleasant of my career. Happy birthday Alberto!

\appendix


\section{The $Z$ Lineshape and Asymmetries}
\label{lineshape}

The $Z$ lineshape measurements determine the cross section 
$e^+ e^- \RA f \bar{f}$ for $f = e, \mu, \tau, s, b, c,$ or hadrons
as a function of $s = E_{CM}^2$. To lowest order,
\beq
\sigma_f(s) \sim \sigma_f \frac{s \Gamma^2_Z}{
  \left(s - M_Z^2\right)^2 + \frac{s^2 \Gamma_Z^2}{M_Z^2}},
\label{sigma}
\eeq
where
significant initial state   radiative corrections are not displayed.

The peak cross section $\sigma_f$ is related to the $Z$ mass and
partial widths by
\beq \sigma_f = \frac{12 \pi}{M_Z^2}  \ \ 
    \frac{\Gamma(e^+ e^-) \Gamma(f \bar{f})}{\Gamma_Z^2}. 
\eeq
The widths are expressed in terms of the effective $Z f \bar{f}$ vector
and axial couplings $\bar{g}_{V,Af} $ by
\beq
\Gamma(f\bar{f}) \sim \frac{C_f G_F M_Z^3}{6 \sqrt{2} \pi}
  \left[ |\bar{g}_{Vf}|^2 + |\bar{g}_{Af}|^2 \right],
\label{width}
\eeq
where $C_\ell = 1$ and $ \ C_q = 3$.
Electroweak radiative corrections are absorbed into the $\bar{g}_{V,Af}$.
There are fermion mass, QED, and QCD corrections to (\ref{width}).

The effective couplings in (\ref{width}) are defined in the SM by
\beq \bar{g}_{Af} = \sqrt{\rho_f} t_{3f}, \ \ \ \
\bar{g}_{Vf} = \sqrt{\rho_f} \left[ t_{3f} - 2 \bar{s}^2_f q_f \right]
\eeq
where $q_f$ is the electric charge and $t_{3f}$ is the weak isospin
of fermion $f$, and $\bar{s}^2_f$ is the effective weak
angle. It is related by ($f$-dependent)
vertex corrections to the on-shell or \msb \ definitions of
\sinn by
\beq \bar{s}^2_f = \kappa_f s^2_W \ \ ({\rm on-shell}) 
=
\hat{\kappa}_f \hat{s}^2_Z  \ \ (\msb). \eeq
$\rho_f-1, \  \kappa_f-1 ,$ and $\hat{\kappa}_f -1 $
are electroweak corrections. For $f=e$ and the known ranges for \mt \ and \mh,
$\bar{s}^2_e \sim \hat{s}^2_Z + 0.00029$.

It is convenient to define the ratios
\beq
R_{q_i}  \equiv \frac{\Gamma(q_i \bar{q}_i)}{\Gamma({\rm had})},\ \  \
R_{\ell_i}  \equiv \frac{\Gamma({\rm had})}{\Gamma(\ell_i \bar{\ell}_i)},
\label{ratios}
\eeq
which isolate the weak vertices (including the effects
of \als \ for $R_{\ell_i}$).
In (\ref{ratios}) $q_i = b, c, s$; $\ell_i = e, \mu, \tau$; and
$\Gamma({\rm had})$ is the width into hadrons. The data are
consistent with lepton universality, i.e., with
$R_e = R_\mu = R_\tau  \equiv R_\ell$.
The partial width into neutrinos or other invisible states
is defined by $
 \Gamma({\rm inv}) = \Gamma_Z - \Gamma({\rm had}) - \sum_i
\Gamma(\ell_i \bar{\ell}_i ),$ where $\Gamma_Z$ is obtained from the
width of the cross section and the others from the peak heights.
This allows the determination of the number of  neutrinos 
by $\Gamma({\rm inv})/\Gamma(\ell \bar{\ell}) 
\equiv N_\nu \Gamma(\nu \bar{\nu})/\Gamma(\ell \bar{\ell})$, 
where $\Gamma(\nu \bar{\nu})$ is the partial
width into a single neutrino flavor.
It has become conventional to work with the parameters
$M_Z, \Gamma_Z, \sigma_{\rm had}, R_\ell, R_b, R_c$, for which the correlations
are relatively small (but still must be included).

The experimenters have generally presented the Born asymmetries, $A^0$,
for which the off-pole, $\gamma$ exchange, $P_{e^-}$, and (small)
box effects have been removed from the data.
Important asymmetries include:
\bqa
{\rm forward-backward:} & & \ \ A^{0f}_{FB} \simeq \frac{3}{4} A_e A_f \nonumber \\
\tau \ \ {\rm polarization:} & & \ \ 
P_\tau^0 = - \frac{A_\tau + A_e \frac{2 z}{1+z^2}}{1
  + A_\tau A_e \frac{2 z}{1+z^2}} \label{asymmetries}
 \\ 
e^- {\rm polarization (SLD):} & & \ \ A^0_{LR} = A_e  \nonumber \\
 {\rm mixed \ \ (SLD):}  & & \ \ 
A^{0FB}_{LR} = \frac{3}{4} A_f \nonumber
\eqa
The LEP experiments also measure a hadronic forward-backward charge asymmetry $Q_{FB}$.
In (\ref{asymmetries}), $A_f$ is defined as the ratio
\beq
 A_f \equiv  \frac{2 \bar{g}_{Vf}  \bar{g}_{Af}}{
\bar{g}_{Vf}^2 + \bar{g}_{Af}^2}
\eeq
for fermion $f$.
The forward-backward asymmetries into leptons allow
another (successful) test of lepton family universality, by
$A^{0e}_{FB} = A^{0\mu}_{FB} = A^{0\tau}_{FB} 
\equiv A^{0\ell}_{FB}$. In the $\tau$ polarization, 
$z = \cos \theta$, where $\theta $ is the  scattering angle.
The SLD polarization asymmetry $A^0_{LR}$ for hadrons (or leptons)
projects out the initial electron couplings. It is especially
sensitive to \sinn because it is linear in the small $\bar{g}_{Ve}$,
while the leptonic $A^{0\ell}_{FB}$ are quadratic. The mixed polarization-FB
asymmetry $A^{0FB}_{LR}$ projects out the final fermion coupling.

\section{Radiative Corrections}
\label{radiativecorr}
The data are sufficiently precise that one must include
high-order radiative corrections, including 
the dominant two-loop electroweak ($\alpha^2 m_t^4, \ \alpha^2 m_t^2$),
dominant 3 loop QCD (and 4 loop estimate),
and dominant 3 loop mixed QCD-EW ($\alpha \alpha_s$ vertex) corrections.

In including EW corrections, one must choose a definition
of the renormalized \sinn. There are several popular choices,
which are equivalent at tree-level, but differ by finite
(\mt \ and \mh \ dependent) terms at higher order. These include
  \begin{itemize}
   \item  On shell: $s^2_W \equiv 1 - \frac{M_W^2}{M_Z^2}$
   \item  $Z$ mass: $s^2_{M_Z} \left(1 - s_{M_Z}^2 \right) \equiv
\frac{\pi \alpha (M_Z)}{\sqrt{2} G_F M_Z^2} $
   \item  \msb: $\hat{s}^2_Z \equiv \frac{\hat{g}'^2 (M_Z)}{\hat{g}'^2 (M_Z) +
\hat{g}^2 (M_Z)}$
   \item   Effective ($Z$-pole):    $\bar{s}^2_f \equiv \frac{1}{4}
  \left( 1 - \frac{\bar{g}_{Vf}}{\bar{g}_{Af}}  \right)$
  \end{itemize}
The first two are {\it defined} in terms of the $Z$ and $W$ masses; the \msb \
from the renormalized couplings $\hat{g}$, $\hat{g}'$; and
the effective from the observed  vertices. Of course, each can be
determined experimentally from any observable, given the appropriate SM expressions.
The $Z$-pole $\bar{s}^2_f$ depends on the fermion $f$ in the final state.
Some of the advantages and drawbacks of each scheme are summarized in
Table~\ref{schemes}.

The expressions for \mw \ and \mz \ in the on-shell and \msb \ schemes
are
\beq 
M_W^2 = \frac{\left( \pi \alpha/\sqrt{2} G_F \right)}{
s_W^2 (1 - \Delta r)} = 
\frac{\left( \pi \alpha/\sqrt{2} G_F \right)}{
 \hat{s}^2_Z (1 - \Delta \hat{r}_W)}
\eeq
and
\beq
M_Z^2 = \frac{M_W^2}{c^2_W} = \frac{M_W^2}{
  \hat{\rho} \hat{c}^2_Z},
\eeq
where the other renormalized parameters are the fine structure constant $\alpha$
(from QED) and the Fermi constant $G_F$, defined in terms of the $\mu$ lifetime.
$\Delta r$, $\Delta \hat{r}_W$, and $\hat{\rho} - 1$ collect
the radiative corrections involving $\mu$ decay, \mw, \mz, and the running
of $\alpha$ up to the $Z$ pole. In \msb, $\Delta \hat{r}_W $
has only weak \mt \ and \mh \ dependence, and is dominated
by the running of $\alpha$, i.e, 
$\Delta \hat{r}_W \sim \Delta \alpha + \cdots  \sim 0.066 + \cdots$.
In contrast, the on-shell $\Delta r$ has an additional
large (quadratic) \mt \ dependence, which results in
a large sensitivity of the observed value of $s_W^2$ to \mt. The 
\msb \ scheme isolates the large effects in the explicit
parameter
$\hat{\rho} \sim 1 + \frac{3 G_F \hat{m}^2_t}{8 \sqrt{2} \pi^2} + \cdots$.
The various definitions are related by (\mt \ and \mh \ dependent)
form factors $\kappa$, e.g., $\bar{s}^2_f = \kappa_f s^2_W = 
\hat{\kappa}_f \hat{s}^2_Z$. For $f=e$ and the experimental \mt, \mh,
one obtains $\bar{s}^2_e \sim \hat{s}^2_Z + 0.00029$.

\begin{table} \centering
 \begin{tabular}{|l|}  \hline \hline
On-shell : $s^2_W = 1 - \frac{M_W^2}{M_Z^2} =  0.22272 \, (38)$ \\
\hline
$+$ most familiar \\
$+$ simple conceptually \\
$-$ large $m_t,\ M_H$ dependence from $Z$-pole observables \\
$-$ depends on SSB mechanism
$-$ awkward for new physics \\ \hline \hline
$Z$-mass : $s^2_{M_Z} = 0.23105\, (8)$ \\ \hline
$+$ most precise (no $m_t,\ M_H$ dependence) \\
$+$ simple conceptually \\
$-$ $m_t,\ M_H$ reenter when predicting other observables \\
$-$ depends on SSB mechanism
$-$ awkward for new physics \\ \hline\hline
$\overline{MS}$ : $\hat{s}^2_Z = 0.23107\, (16)$ \\ \hline
$+$ based on coupling constants \\
$+$ convenient for and minimizes mixed QCD-EW corrections \\
$+$ convenient for GUTs \\
$+$ usually insensitive to new physics \\
$+$ $Z$ asymmetries $\sim$ independent of $m_t,\ M_H$ \\
$-$ theorists definition; not simple conceptually \\
$-$ usually determined by global fit \\
$-$ some sensitivity to $m_t,\ M_H$ \\
$-$ variant forms ($m_t$ cannot be decoupled in all processes; \\
\ \ \ $\hat{s}^2_{ND}$ larger by $0.0001 - 0.0002$) \\ \hline \hline
effective : $\bar{s}^2_{\ell} = 0.23136 \, (15)$ \\ \hline
$+$ simple \\
$+$ $Z$ asymmetry independent of $m_t$ \\
$+$ $Z$ widths: $m_t$ in $\rho_f$ only \\
$-$ phenomenological; exact definition in computer code \\
$-$ different for each $f$ \\
$-$ hard to relate to non $Z$-pole observables \\ \hline \hline
\end{tabular}
\caption{Principle definitions of the
renormalized \sinn and their features.}
\label{schemes}
\end{table}

The \msb \ weak angle $\hat{s}^2_Z$ can be obtained cleanly from
the weak asymmetries. Comparison with \mz \ and \mw \ is important
for constraining \mh \ and new physics. The largest theory uncertainty in the
$M_Z-\hat{s}^2_Z$ relation is the hadronic
contribution to the running of $\alpha$ from
its precisely known value $\alpha^{-1} \sim 137.036 $ at low energies,
to the electroweak scale, where one expects
$\alpha^{-1}(M_Z) \sim \hat{\alpha}^{-1}(M_Z) + 0.99 \sim 129$.
($\hat{\alpha}$ refers to the \msb \ scheme.)
There is a related uncertainty in the hadronic contribution to the
anomalous magnetic moment of the muon.
More explicitly, one can define $\Delta \alpha$ by
\beq
\alpha(M_Z^2) = \frac{\alpha}{1-\Delta \alpha}.
\eeq
Then,
\beq \Delta \alpha = \Delta \alpha_\ell + \Delta \alpha_t +
\Delta \alpha^{(5)}_{\rm had}
\sim  0.031497 - 0.000070   + \Delta \alpha^{(5)}_{\rm had}. \eeq
The  leptonic and $t$ loops are reliably calculated
in perturbation theory, but not $ \Delta \alpha^{(5)}_{\rm had}$ from the
lighter quarks. $ \Delta \alpha^{(5)}_{\rm had}$ can be expressed
by a dispersion integral involving $R_{\rm had}$ (the cross section
for $e^+ e^ - \RA$ hadrons relative to  $e^ + e^- \RA \mu^+ \mu^-$).
Until recently, most calculations  were data driven, using experimental values for
$R_{\rm had}$ up to CM energies $\sim$ 40 GeV, with perturbative
QCD (PQCD) at higher energies. However, there are  significant experimental uncertainties
(and some discrepancies) in the low energy data. A number of recent
studies have argued that one could reliably use a combination
of theoretical estimates using PQCD and such non-perturbative techniques as
sum rules and operator product expansions down to $\sim$ 2 GeV,
leading to a different (usually lower) central value, and lower
uncertainties. New BES-II data from Beijing have
reduced the central value and uncertainty in the data driven approach,
leading to a partial convergence of the two techniques. Recent values are listed
in Table~\ref{alhadvalues}. One can also determine $ \Delta \alpha^{(5)}_{\rm had}$
directly from the precision fits (Section~\ref{globalfits}), in agreement
with these  estimates but with a larger uncertainty.

There have been experimental observations of the running of $\alpha$
by TOPAZ ($e^+ e^- \mu^+ \mu^-$);
 VENUS, L3 (Bhabha), and OPAL (high $Q^2$). While not sufficiently precise
to determine $ \Delta \alpha^{(5)}_{\rm had}$, these are interesting
confirmations of QED.
\begin{table} \centering
\begin{tabular}{|l|c|l|}
\hline
Author(s)                 & Result              & Comment                                     \\
\hline
Martin \& Zeppenfeld      & $0.02744 \pm 0.00036$ & PQCD for $\sqrt{s} >  3$~GeV              \\
{\it Eidelman \& Jegerlehner}   & 
{$0.02803 \pm 0.00065$ }& PQCD for $\sqrt{s} > 40$~GeV              \\
Geshkenbein \& Morgunov   & $0.02780 \pm 0.00006$ & ${\cal O}(\alpha_s)$ resonance model      \\
Burkhardt \& Pietrzyk     & $0.0280  \pm 0.0007 $ & PQCD for $\sqrt{s} > 40$~GeV              \\
Swartz                    & $0.02754 \pm 0.00046$ & use of fitting function                   \\
Alemany, Davier, H\"ocker & $0.02816 \pm 0.00062$ & includes $\tau$ decay data                \\
Krasnikov \& Rodenberg    & $0.02737 \pm 0.00039$ & PQCD for $\sqrt{s} > 2.3$~GeV             \\
Davier \& H\"ocker        & $0.02784 \pm 0.00022$ & PQCD for $\sqrt{s} > 1.8$~GeV             \\
K\"uhn \& Steinhauser     & $0.02778 \pm 0.00016$ & complete ${\cal O}(\alpha_s^2)$           \\
{\it Erler }                    &
{ $0.02779 \pm 0.00020$ }& converted from $\overline{\rm MS}$ scheme \\
Davier \& H\"ocker        & $0.02770 \pm 0.00015$ & use of QCD sum rules                      \\
Groote {\it et al.}       & $0.02787 \pm 0.00032$ & use of QCD sum rules                      \\
Jegerlehner               & $0.02778 \pm 0.00024$ & converted from MOM \\
Martin, Outhwaite, Ryskin & $0.02741 \pm 0.00019$ & includes new BES data                     \\
{\it Pietrzyk  }                & $0.02755 \pm 0.00046$ & details not
published                     \\
\hline
\end{tabular}
\caption{Recent evaluations of  $\Delta \alpha_{\rm had}^{(5)}(M_Z)$
 (adjusted to  $\alpha_s(M_Z) = 0.120$).}
\label{alhadvalues}
\end{table}


\newpage
\begin{figure}[h]
\postscript{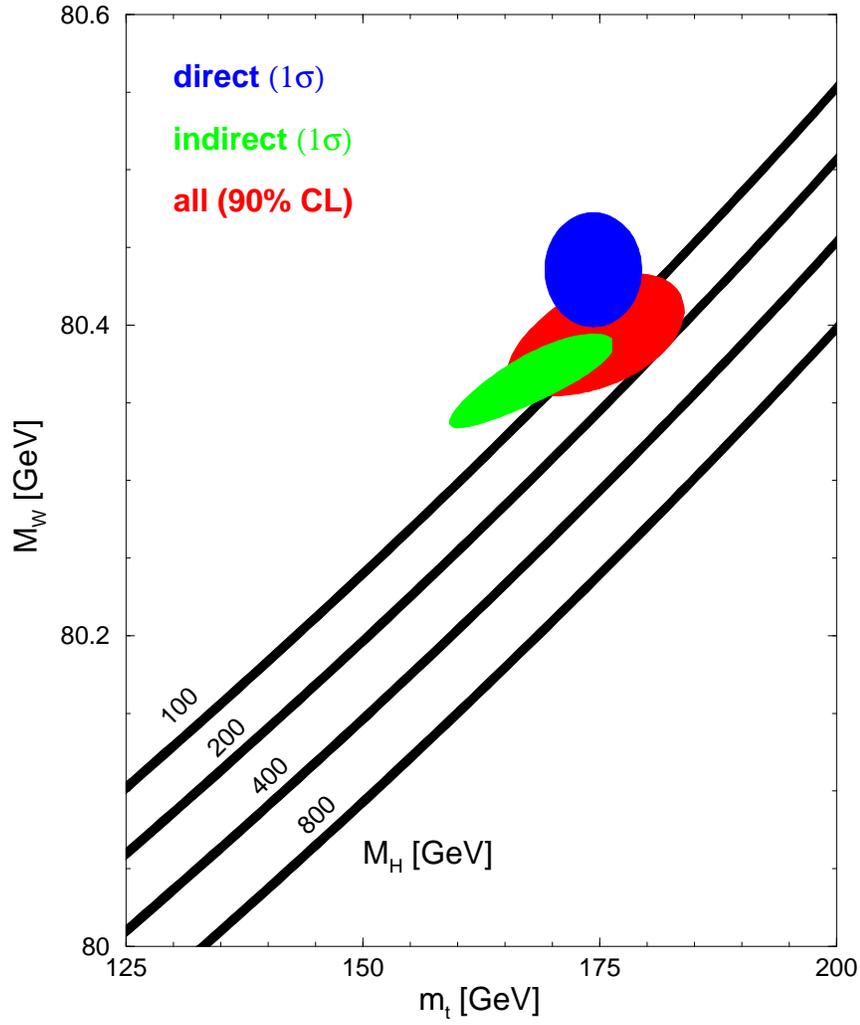}{0.8}
\caption{Allowed regions in $M_W$ vs \mt \ from direct, indirect,
and combined data, compared with the standard model expectations as a function of
\mh. From~\cite{erler}.}
\end{figure}

\begin{figure}[h]
\postscript{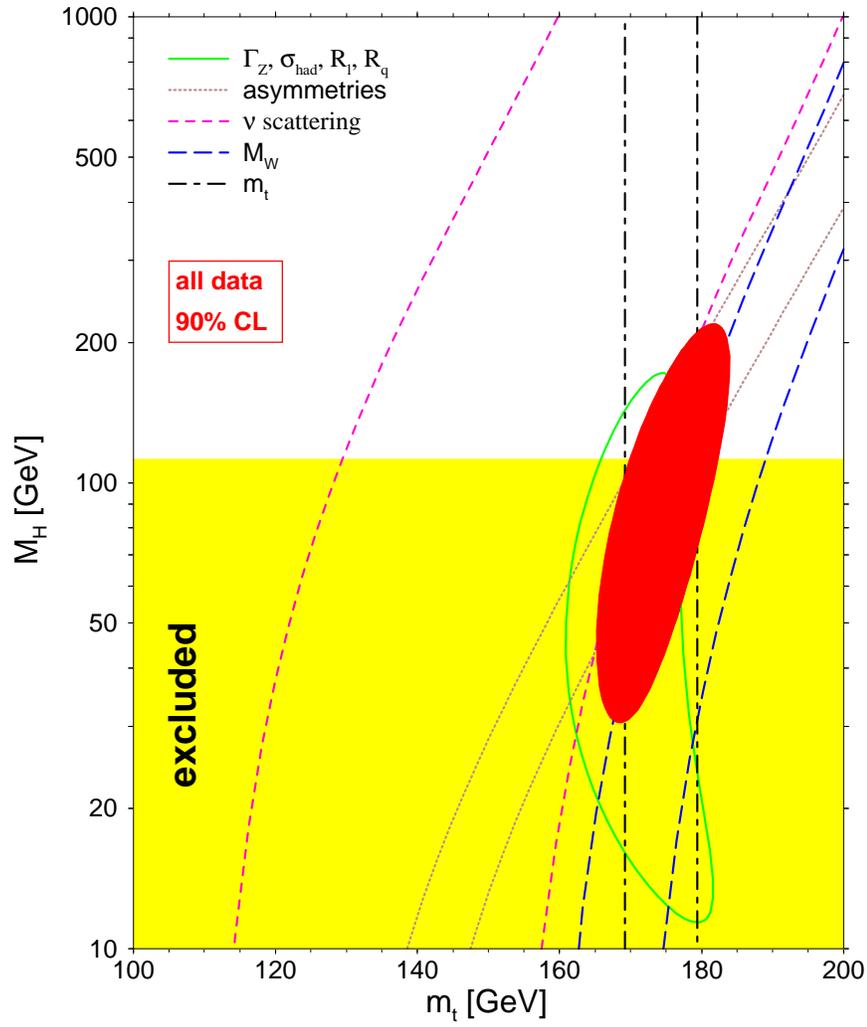}{0.8}
\caption{Allowed regions in \mh \  vs \mt \ from precision data,
compared with the direct exclusion limits from LEP~2.
From~\cite{erler}.}
\end{figure}

\begin{figure}[h]
\postscript{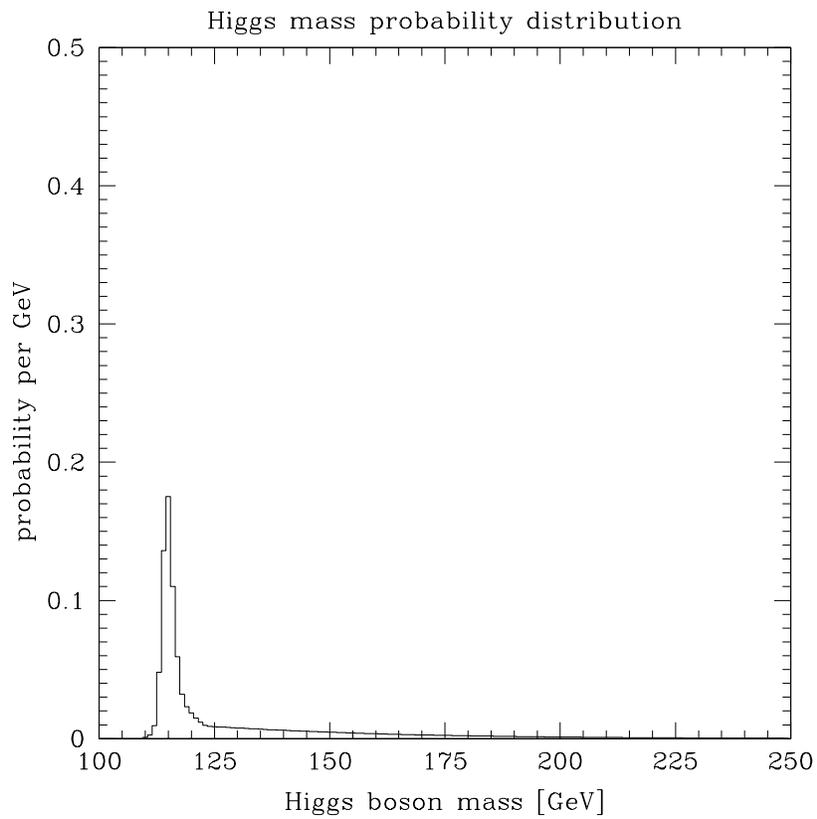}{0.8}
\caption{Probability density for \mh, including direct LEP~2 data and indirect
constraints. From~\cite{erlerhiggs}.}
\end{figure}

\begin{figure}[h]
\postscript{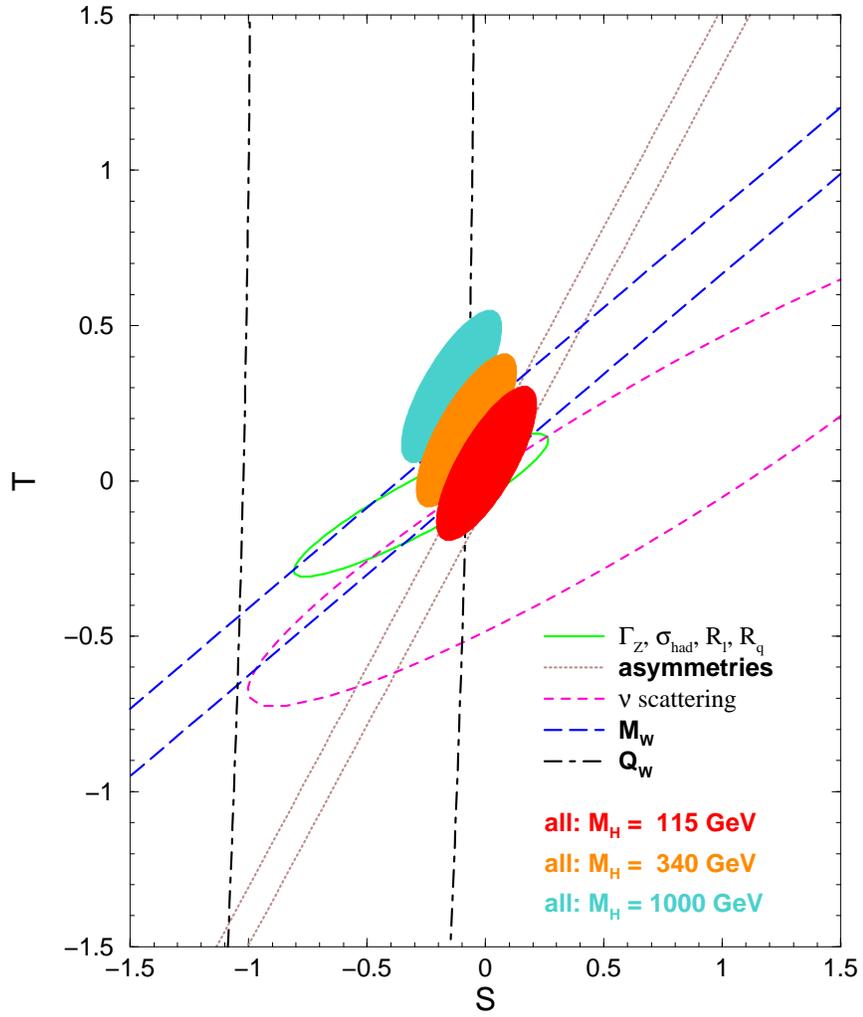}{0.8}
\caption{Allowed regions in $S$ vs $T$.
From~\cite{erler}.}
\end{figure}

\begin{figure}[h]
\postscript{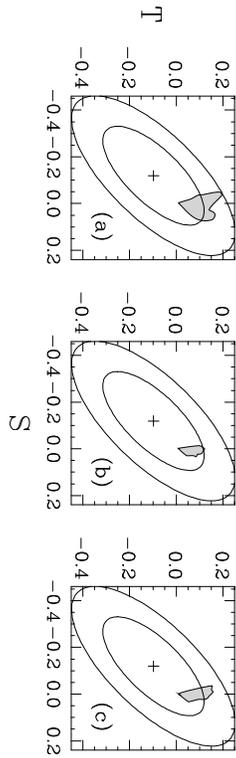}{0.9}
\caption{Predicted values of $S$ and $T$ for three models of supersymmetry
breaking, compared with the experimentally allowed regions. Updated from~\cite{susy}.}
\end{figure}

\begin{figure}[h]
\postscript{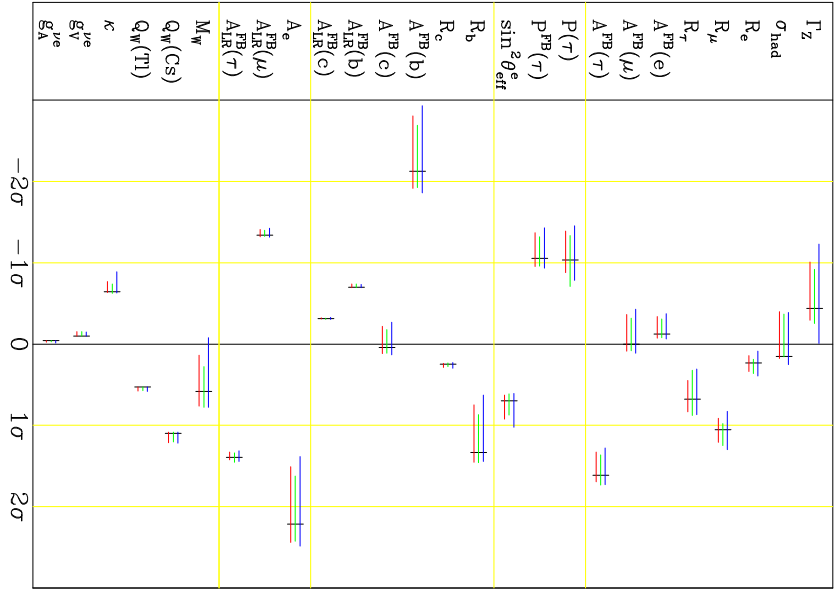}{0.8}
\caption{Changes in pulls of observables in three models of supersymmetry breaking, compared
with the standard model.
Updated from~\cite{susy}.}
\end{figure}

\begin{figure}[h]
\postscript{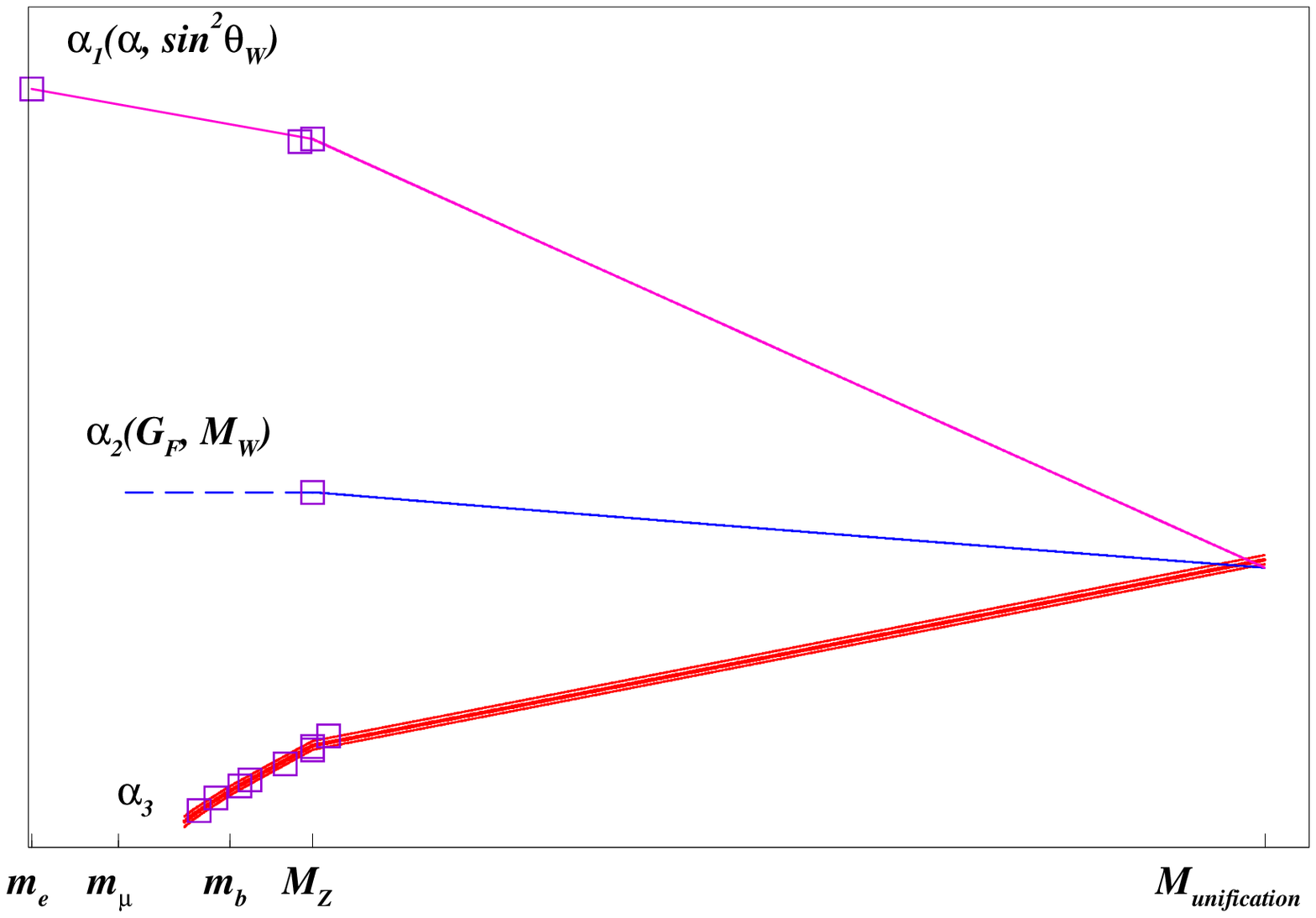}{0.8}
\caption{Running inverse gauge couplings in the MSSM. Experimental values at low
energies are also shown. From~\cite{polonsky}.}
\end{figure}


\begin{thebibliography}{99}
\bibitem{general} For complete references, see
J. Erler and P. Langacker, 
{\it Electroweak Model and Constraints on New Physics},
in {\it Review of Particle Physics}, 
D.~E.~Groom {\it et al.},
Eur.\ Phys.\ J.\ {\bf C15}, 1 (2000);
{\it Precision Tests of the Standard Electroweak Model},
ed. P. Langacker
(Singapore, World, 1995);
P.~Langacker, M.~Luo and A.~K.~Mann,
Rev.\ Mod.\ Phys.\ {\bf 64}, 87 (1992).
\bibitem{kim}  J.~E.~Kim {\it et al.},
Rev.\ Mod.\ Phys.\ {\bf 53}, 211 (1981).
\bibitem{amaldi} U.~Amaldi {\it et al.},
Phys.\ Rev.\ {\bf D 36}, 1385 (1987).
\bibitem{costa} G.~Costa  {\it et al.},
Nucl.\ Phys.\ {\bf B297}, 244 (1988).
\bibitem{unique} P.~Langacker,
Comments Nucl.\ Part.\ Phys.\ {\bf 19}, 1 (1989).
\bibitem{marciano} W. Marciano, these proceedings.
\bibitem{erler} Updated from~\cite{general}.
\bibitem{LEPEWWG} LEP Collaborations,
hep-ex/0101027. Figures and other materials are available at
 http://www.cern.ch/LEPEWWG/. \hfill
\bibitem{bardin} D. Bardin, these proceedings.
\bibitem{hollik}  W. Hollik, these proceedings.
\bibitem{passarino} G. Passarino,  these proceedings.
\bibitem{kinoshita}  T. Kinoshita, these proceedings.
\bibitem{jegerlehner}  F. Jegerlehner, these proceedings.
\bibitem{giga}
J.~Erler, S.~Heinemeyer, W.~Hollik, G.~Weiglein and P.~M.~Zerwas,
Phys.\ Lett.\ {\bf B486}, 125 (2000).
\bibitem{GAPP} J.~Erler,
hep-ph/0005084.
GAPP, as well as fit results,  are available at
www.physics.upenn.edu/$\sim$erler/electroweak/
\bibitem{alhad}  J. Erler, Phys. Rev. D59, 054008 (1999).
\bibitem{erlerhiggs} J.~Erler,
hep-ph/0010153 and these proceedings.
\bibitem{degrassi} G. Degrassi, these proceedings.
\bibitem{gurtu}  A. Gurtu, {\it Precision Tests of the Electroweak Gauge Theory},
ICHEP  2000, Osaka, August 2000.
\bibitem{ej} S. Eidelman and F. Jegerlehner, Z. Phys. C67, 585 (1995).
\bibitem{Pietrzyk} B. Pietrzyk, {\it The Global Fit to Electroweak Data},
ICHEP  2000.
\bibitem{susy} J.~Erler and D.~M.~Pierce,
Nucl.\ Phys.\ {\bf B526}, 53 (1998).
\bibitem{polonsky} 
P.~Langacker and N.~Polonsky,
Phys.\ Rev.\ {\bf D 52}, 3081 (1995).
\bibitem{ack}
This work was supported by the U.S. Department
of Energy grant DOE-EY-76-02-3071
\end{thebibliography}
\end{document}